\documentclass[a4paper]{jpconf}
\usepackage{graphicx}
\begin{document}
\title{Physics of the Muon Spectrometer of the ALICE Experiment
\footnote{Talk presented in the ICPAQGP Conference, February 8-12, 2005, Salt Lake City, Kolkata, India.
Web page of the conference : http://www.veccal.ernet.in/$\sim$icpaqgp/}}

\author{Gin\'es MARTINEZ \footnote{martinez@in2p3.fr}\\
for the ALICE Collaboration}

\address{SUBATECH (EMN-UN-IN2P3), 4 rue Alfred Kastler BP20722, 44307 Nantes Cedex 3}

\begin{abstract}

The main goal of the ALICE Muon spectrometer experiment is the measurement
of heavy quark production in p+p, p+A and A+A collisions at LHC energies, via the muonic channel.
Physics motivations and expected performances have been presented in this talk.
\end{abstract}

\section{Physics motivations}
Heavy ion collisions (HIC) at relativistic energies are a privileged tool for creating very hot and dense matter in a laboratory.
In particular, lattice chromo-dynamics (lQCD) predicts a cross-over toward a new state of matter called Quark Gluon Plasma (QGP) 
at a temperature $\sim~170$ MeV for vanishing chemical potential $\mu_B$ \cite{Kars01,Ejir04,Gava05}.
Heavy ion collisions allow to experimentally study the properties of this new state of matter.
This experimental program started in the mid 80's with fixed target heavy ion experiments at the AGS and 
SPS \cite{Satz02} 
and continued with the physics program developed at the RHIC collider (BNL) \cite{Hemm04}.
Heavy ion collisions at the future Large Hadron Collider (LHC) at CERN 
will open new experimental insights in the study of hadronic matter at high temperature.
The ALICE experiment will be the only experiment at the LHC devoted to heavy ion physics, 
whereas the ATLAS and the CMS experiments plan to develop a heavy-ion program \cite{Schu05,Taka04} in parallel with their
main physics goals.

The LHC collider will provide lead (argon) high luminosity beams $<\mathcal{L}>=5 \cdot 10^{26}$cm$^{-2}$s$^{-1}$  ($<\mathcal{L}>=5 \cdot 10^{28}$cm$^{-2}$s$^{-1}$)
at $\sqrt{s}=$5.5A TeV ($\sqrt{s}=6.3$A TeV).
In addition, the LHC will deliver proton beams $<\mathcal{L}>=3\cdot 10^{30}$cm$^{-2}$s$^{-1}$ at  $\sqrt{s}=$14 TeV and 
d+Pb beams $<\mathcal{L}>=8\cdot 10^{28}$cm$^{-2}$s$^{-1}$ at $\sqrt{s}\sim$6.2A TeV,
providing a solid baseline for the study of medium effects in HIC \cite{AlicePPR1}
\footnote{
It should be noted that d+Pb and Ar+Ar beams will require an upgrade of the LHC collider. 
Reported luminosities are very preliminary \protect\cite{Mors05}.}.
At such ultra-relativistic energies new phenomena emerge, improving the experimental conditions for studying 
the hadronic matter in HIC: 
\begin{itemize}
\item \textbf{Initial conditions.} The initial conditions will be under control by the gluon saturation scenario.
At these energies the initial nucleus-nucleus interaction can be viewed as \emph{weak} 
interactions of a huge number of \emph{small x} gluons which are freed in the beginning 
of the collision leading to the formation of
a big gluonic ball \cite{Lerr01,Ianc03}.
Up to 8000 gluons in the early stage of the collisions are predicted \cite{Venu01}. 
Such a system will rapidly evolve toward a equilibrium \cite{Baie01}. 
Most of these processes (as secondary interaction of minijets) leading to thermalization will be 
governed by hard processes ($\alpha_s<1$).  
\item \textbf{Equilibrated matter.} After equilibration of the initial gluonic ball, a hot and long-lived 
hadronic matter will be formed (hotter and longer-lived than at lower energy HIC). 
The increase of the beam energy will favor the creation of vanishing baryonic potential  
hadronic matter with a temperature around 0.5-1 GeV,
well above the critical temperature predicted by lQCD.
\item \textbf{Observables.} The experience acquired during the last 20 years of Heavy-Ion Physics in the relativistic regime has 
shown the necessity to measure most of the probes of the reaction dynamics, 
from hard processes and QGP formation observables, until the freeze-out of the expanding hadron gas.
A coherent explanation of the full set of observables will be the only way to 
study the properties of the ephemeral QGP. 
Final states probes like particle multiplicities, hadron $p_T$ distributions, particle ratios, strangeness production,
azimuthal asymmetries, etc ...  
will shed light on the condition of the phase transition and the dynamical evolution of the expanding hadron gas. 
Penetrating probes, as real and virtual photon production, charmonium production, and light vector meson properties  
will provide informations about the QGP formation and properties.
Moreover, the study of the QGP at LHC energies will be enriched by exploiting new  
probes which can be efficiently studied in this energy regime:
\begin{enumerate}
\item Hard processes leading to very fast leading partons will provide information about the QGP through the interaction
of jets with the surrounding partonic matter \cite{Wied00}. 
Huge parton energy losses (around 1-3 GeV/fm) in the QGP will modify the hadronization process 
of these very energetic jets, inducing a jet-quenching and a suppression of high p$_T$ hadron production. 
At the LHC, jet production cross-section in the very high $p_T$ range (p$_T>$50 GeV) will increase by more of 4 
orders of magnitude with respect to RHIC beam energies;
\item The Debye screening of bottomonium states in QGP will be studied. 
Suppression of the Upsilon family production will be exploited due to its sensitivity to the
density of color charges in the medium \cite{Satz86};
\item Open charm and open beauty production will be exploited as new probes of the strong 
interacting system of partons \cite{Croc05,Wong98}. Open charm (beauty) production cross-section will increase by a factor 10 (100)
with respect to RHIC energies; 
\item The production of massive electroweak bosons (W$^{+-}$ and Z) will open the possibility to check the validity of 
the glauber scaling in HIC. These bosons do not interact with the surrounding medium and they are produced in hard parton collisions;
\item Finally, the huge particle multiplicity at the LHC 
will make possible to measure a large number of observables on an event-by-event basis, 
increasing the sensitivity to non-statistical fluctuations predicted to occur in a phase transition scenario.
\end{enumerate}
\end{itemize}

\section{Heavy Quark Production at the LHC}
Heavy quarks (charm and beauty quarks) in HIC are produced in the first stages of the collisions and then they coexist with the surrounding medium due to their long life-time. 
Production rates, transverse momentum ($p_T$) and rapidity ($y$) distributions, quarkonia production rates, etc ... 
will allow for probing the properties of the medium, as the QGP.

\subsection{Initial hard production}
Charm quarks will be copiously produced in HIC at LHC energies: more than 100 
$c\bar{c}$ pairs per central Pb+Pb collision and around 5 $b\bar{b}$ pairs per collision are expected \cite{Yell03}.
Production of heavy quarks will be dominated by prompt parton-parton scattering. 
Open heavy flavor production will be dominated by gluon fusion. Next leading-order (NLO) diagrams noticeably contribute to the heavy quark 
production cross-section \cite{Naso89} although the LO one particle differential cross-section shapes are not appreciably modified 
by the NLO contribution.
It has been observed that pQCD calculations underestimate the measured differential cross-sections in $p\bar{p}$ collisions at Tevatron energies, 
although they are compatible within uncertainties \cite{CDF03}. 
All those experimental data concerns high p$_T$ production (larger than 5 GeV/c).
At LHC energies, heavy quark production will explore a $x$ and transfered momentum $Q^2$ domain which has not been well studied.
For instance in rapidities $2.5<|y|<4$ at LHC charm production will be sensible to $x\sim10^{-4}$ and $Q^2\sim10$ GeV$^2$.
In the case of HIC, one should expect  gluon shadowing which will modify the heavy quark production based on glauber scaling.
It is expected a heavy quark yield suppression of about 20\% (10\%) for charm (beauty) production in p+Pb collisions at LHC energies \cite{Rauf04} 
due to gluon shadowing. 
 
About 1\% of the heavy quark pairs will end in the formation of a colorless bound state: quarkonium. 
The process of direct formation of quarkonia from the initially produced $Q\bar{Q}$ pair is a non-perturbative process.
Models like Color Evaporation Model (CEM) or non-relativistic QCD \cite{Yell03} have been successfully tested in $pp$ and $p\bar{p}$ collisions.
In HIC direct production of quarkonia will take place inside the impinging nuclei, leading to an initial nuclear absorption of quarkonia 
(normal suppression in HIC).
The nuclear absorption increases with the enhancement of the relative rapidity between the quarkonia and the impinging nuclei.
For this reason, nuclear absorption becomes higher at LHC energies, a factor 2 higher than at SPS energies. 
The nuclear absorption exhibits a relatively flat distribution as a function of the HIC centrality for central and semi-central collisions 
and becomes maximal for central Pb+Pb collisions at the LHC, where the survival probability of the charmonium is $\sim$ 25\% (50\% for bottomonium) \cite{Yell03}. 

\subsection{Pre-equilibrium production}
Pre-equilibrium stage at LHC energies will be dominated by secondary mini-jet interactions
which could noticeably contribute to the total production cross-section of charm quark pairs \cite{Mull92}.
One could expect up to 200 $c\bar{c}$  in central Pb+Pb collisions taking into account the production of heavy quarks during the thermalization phase.

\subsection{Heavy quarks in hot and dense medium} 
Heavy quarks will be then embedded in a matter mainly formed by gluons and light quarks ($u,d,s$).
It is a open question how heavy quarks will behave in such a medium.
Will heavy quarks thermalize? Will heavy quarks develop collective motion? 
Some models assume that heavy quarks behave like \emph{Brownian} particles (m$_Q>$m$_q$) \cite{Goss04} and
their transverse momentum (p$_T$) and rapidity distributions, collective motion ... 
will probe the properties of the QGP.
Although pQCD heavy quark interaction cross-sections with the medium seems too small for reaching full thermalization of heavy quarks, 
non-perturbative effects in QGP could noticeably increase the heavy-quark cross-section, leading to their full thermalization \cite{Hess05}.
Other models do not address the problem of thermalization and directly assume statistical coalescence of heavy quarks \cite{Brau00,Andr03,Gran04,Thew01}. 

The study of the heavy quark bound states will allow for probing the medium via 
the Debye screening \cite{Gava05, Datt04}, the gluon dissociation of quarkonia \cite{Gran04}, 
kinetic recombination of heavy quarks \cite{Thew01} and/or their statistical hadronization \cite{Brau00,Andr03}.
It is expected that charmonium production in HIC at LHC will be dominated by recombination of $c\bar{c}$ pairs in the QGP or at the hadronization. 
Due to the huge production rate of charmed quarks at LHC, these models predict a spectacular enhancement of the charmonium yield
in central Pb+Pb collisions, more than a factor 10.
Based on these models, charmonium yield will exhibit centrality dependence proportional to $<N_{c\bar{c}}>^2$ \cite{Andr03,Gran04,Thew01}, 
which will be a very clear signature of QGP.
In the case of charmonium, a non-negligible production could occur after hadronization of the QGP due to $D\bar{D}$ annihilation \cite{Ko98,Brau00}.
For the bottomonium production at the LHC, the recombination mechanism will play a less important role due to the lower bottom pair production rate. 
Recombination mechanism would compete with the suppression mechanism due to Debye screening and/or gluon dissociation.
Charm measurements at RHIC would help to predict what should occur with beauty at LHC
since production rates per central HIC are similar.


\subsection{Secondary production}
There will be a production of $J/\psi$ due to charmonium decay modes of the B mesons.
At LHC energies the relative contribution of $J/\psi$ from B-meson decay should be about 25\% \cite{Croc05}.
Upsilon family production will be increased by radiative decay of $\chi_b$ quarkonia.
About 60\% of the total Upsilon family production will originate from these resonances.

\section{Production of electroweak bosons}
Electroweak bosons ($W^{+,-}$, $Z$) will be abundantly produced at LHC and this will allow for several precision measurements, 
for instance high precision measurement of the W mass \cite{Atla99}. 
However, proton luminosities in ALICE experimental region will be much lower ($<\mathcal{L}>=3\cdot 10^{30}$cm$^{-2}$s$^{-1}$) 
than in other LHC experiments, due to ALICE detector performances.
At leading-order, W bosons are produced in quark - antiquark collisions: $u\bar{d}\rightarrow W^+$, $c\bar{s}\rightarrow W^+$ etc ...
The LO production cross-section of $W^{+-}$ obtained from Pythia \cite{Pythia} amounts to $\sim$17 nb in p+p collisions at 14 TeV \cite{Cone05}, 
leading to a production of half a millions W bosons decaying in the muonic channel in one year of data taken in the ALICE interaction point.
At these energies, contributions to $W$ production from higher order diagrams should be about 10\% \cite{Frix05}.
In proton+proton collisions, there will be more $u$ than $d$ valence quarks, leading to a enhancement of $W^+$ with respect $W^-$ 
which is more pronounced at high rapidities where the W-boson production is dominated by valence-quarks.
At mid-rapidity the production  of $W^+$ and $W^-$ is almost symmetric.

In HIC, W bosons production will not be affected by the medium effects since W bosons will leave the hot and 
dense bulk of matter without any secondary interaction. In this respect, W bosons production will be a very powerful tool for checking 
the validity of the glauber scaling hypothesis at these energies.
Since shadowing could be different in quarks and gluons, measurements of W production in
d+Pb (or p+Pb)  will be necessary.
W-boson production could provide a powerful normalization of very high $p_T$ ($p_T\sim$30-40 GeV/c) beauty and charm production
where a suppression could be expected due to jet quenching of heavy quarks.
The asymmetry between $W^+$ and $W^-$ in HIC will be given by the isospin of the nuclei and the elementary asymmetry in p+p, n+n, p+n and n+p 
collisions \cite{Cone05}.

\section{Muon Spectrometer}
In the framework of the ALICE physics program \cite{Schu05},
the goal of the Muon spectrometer of ALICE \cite{Finc05} is the study of open heavy flavor production and quarkonia production 
($J/\psi$, $\psi'$ and $\Upsilon$(1$S$), $\Upsilon$(2$S$) and $\Upsilon$(3$S$)) via the muonic channel.
For HIC the dependence with the collision centrality and with the reaction plane will also be studied.
The study of electroweak bosons is also foreseen and preliminary  studies are under progress \cite{Cone05}.

\begin{figure}
\begin{center}
\includegraphics[width=100mm,clip]{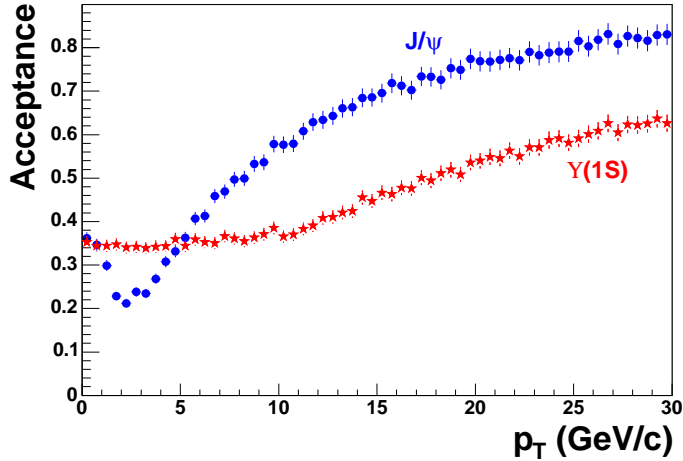}
\end{center}
\caption{
{\small Acceptance of the Muon spectrometer as a function of the transverse momentum for $J/\psi$ and 
$\Upsilon(1S)$ in the rapidity range $-4.0<y<-2.5$, 
via their muon pair decay and with a muon low p$_T$ threshold equal to 1 GeV/c.}
\label{fig:acceptance}}

\end{figure}

The main experimental requirement is to measure the quarkonia production 
in central Pb+Pb collisions at LHC energies, down to very low p$_T$, since
low p$_T$ quarkonia will be sensitive to medium effects like heavy-quark potential screening.
Since muons are passively identified by the absorber technique,
a Lorentz boost is needed to be able to measure quarkonia at low p$_T$.
On the other hand, the Muon spectrometer has to be as close as possible to the physics of the QGP 
which occurs in the mid-rapidity region.
As a compromise, the muon spectrometer allows for measuring muons and quarkonia production in an intermediate 
rapidity range $-4.0<y<-2.5$.
The acceptance plot in this rapidity range for $J/\psi$ and $\Upsilon(1S)$ mesons decaying into muon pairs 
is presented in Fig. \ref{fig:acceptance}.
The muon spectrometer will be a unique apparatus at the LHC to measure charmonia production down to p$_T=0$
and will cover a rapidity range which completes the one measured by the ALICE central barrel, and by the CMS and ATLAS experiments.
The other important experimental requirement for the muon spectrometer is to be able to disentangle the different 
resonances of the $\Upsilon$ family. In particular the separation between the resonance $\Upsilon(2S)$
and $\Upsilon(3S)$ ($\Delta$M $\sim$ 300 MeV/c$^2$) imposes to the spectrometer 
an  invariant mass resolution about 100 MeV/c$^2$ in the $\Upsilon$ mass region.

In order to reach these requirements, the muon spectrometer is located downstream (side C of point 2 at LHC) 
of the ALICE detector covering the angular range $171.^\circ<\theta<178.^\circ$, (-4.0$<y<$-2.5)
consisting of 3 absorbers, a muon magnet, a trigger system and a tracking system.
A more detailed description of the Muon spectrometer is given in these proceedings \cite{Finc05}.

\section{Physics program in the Muon spectrometer of ALICE}
The Muon physics program is focused on the measurement of heavy flavor production in p+p, p+A like and A+A collisions at LHC energies.
Many studies have been undertaken \cite{Smbat04, Guer04, Dumo05} mainly in the framework of the Technical Design Report \cite{Muon99} and 
the Physics Performance Report of ALICE \cite{AlicePPR2} and other studies are still in progress.

\subsection{Quarkonia measurements}
\begin{figure}
\begin{center}
\includegraphics[height=7cm,clip]{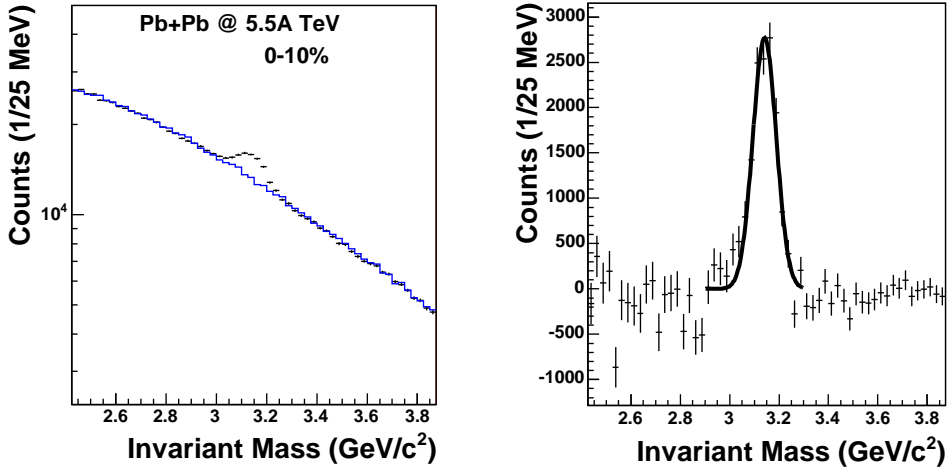}
\caption{\small{Invariant mass distribution of unlike-sign muon pairs in the Muon spectrometer for Pb+Pb collisions at 
LHC energies in the most central collisions (0-10\%).}}
\label{fig:JpsiCentral}
\end{center}
\end{figure}

Invariant mass analysis of muon  pairs will provide a direct measurement of the quarkonia production:
$J/\psi$, $\psi'$,  $\Upsilon(1S)$, $\Upsilon(2S)$, $\Upsilon(3S)$. 
Quarkonia measurement in central Pb+Pb collisions will be very challenging due to the huge low-energy 
background \cite{Muon99,Zint03} and the large muon combinatorial background 
(see in Tab. \ref{smbatTable} the quarkonia signal to background ratio and the significance from reference \cite{Smbat04}). 
In the case of $J/\psi$, a ratio signal to background (S/B) 
in the order of $\sim$ 0.15 is expected in most central (0-10\%) Pb+Pb collisions at 5.5A TeV (see Fig. \ref{fig:JpsiCentral}).
During the first Pb+Pb run with nominal luminosity at the LHC \cite{AlicePPR2}
(this corresponds to a period of 10$^6$ s and  an average  luminosity of $5 \cdot 10^{26}$ cm$^{-2}$ s$^{-1}$) 
we expect to measure about $\sim 6\cdot 10^5$ of $J/\psi$ 
and $\sim 6500$ $\Upsilon(1S)$ \footnote{Glauber scaling and shadowing have been taken into account for this estimation}. 
Those numbers would fluctuate depending on the physics of the heavy quark production and quarkonia production in heavy ion collisions: 
from total suppression due to color screening, to enhancement due to recombination.
Measurement of the ratio of  $\Upsilon(2S)$/$\Upsilon(1S)$  will provide a very powerful experimental observable to constraint 
the different models on quarkonia suppression in the QGP \cite{Gun97, Dumo05}.
The measurement of $J/\psi$ elliptic flow will also be possible and the study of very high p$_T$ $J/\psi$ will represent a very promising hard probe.
The increase of luminosity for the Ar+Ar will allow for cumulating about a factor 5 more quarkonia statistics than in Pb+Pb LHC run 
\begin{table}[h]
\begin{center}
\begin{tabular}{lllllll}
\hline
& $b$ (fm) & 0-3 & 3-6 & 6-9 & 9-12 & 12-16  \\
\noalign{\smallskip}\hline\noalign{\smallskip}
$J/\psi$ & S/B                      & 0.167 & 0.214 & 0.425 & 1.237 & 6.243 \\
         & ${\rm S}/\sqrt{\rm S+B}$ & 111.3 & 180.4 & 213.8 & 193.4 & 94.95 \\
\noalign{\smallskip}\hline\noalign{\smallskip}
$\psi^\prime$ & S/B                 & 0.009 & 0.011 & 0.021 & 0.063 & 0.273 \\
         & ${\rm S}/\sqrt{\rm S+B}$ & 4.185 & 6.902 & 8.604 & 9.641 & 7.171 \\ 
\noalign{\smallskip}\hline\noalign{\smallskip}
$\Upsilon$ & S/B                    & 2.084 & 2.732 & 4.31  & 7.977 & 12.01 \\
           & ${\rm S}/\sqrt{\rm S+B}$ & 27.39 & 41.71 & 40.03 & 27.16 & 10.42\\
\noalign{\smallskip}\hline\noalign{\smallskip}
$\Upsilon^\prime$ & S/B           & 0.807 & 1.043 & 1.661 & 2.871 & 4.319 \\
       & ${\rm S}/\sqrt{\rm S+B}$ & 11.68 & 18.26 & 18.48 & 13.02 & 5.077 \\ 
\noalign{\smallskip}\hline\noalign{\smallskip}
$\Upsilon^{\prime\prime}$ & S/B & 0.566 & 0.722 & 1.18  & 1.936 & 3.024 \\
& ${\rm S}/\sqrt{\rm S+B}$ & 7.951 & 12.55 & 13    & 9.274 & 3.73 \\       
\hline
\end{tabular}
\end{center}
\caption{{\small Preliminary quarkonia signal over background (S/B) and significance
(${\rm S}/\sqrt{\rm S+B}$) for quarkonium resonances measured versus centrality
in the ALICE forward muon spectrometer~\protect\cite{Smbat04}.
The input cross-sections are taken from~\protect\cite{Yell03}.
Shadowing is taken into account.
Any other suppression or enhancement effects are not included.
The numbers correspond to one month of Pb-Pb data taking and are extracted with a 2$\sigma$ mass cut.
\label{smbatTable}   }}
\end{table}

In the case of proton-proton run at 14 TeV \cite{AlicePPR2}
(this corresponds to a period of 10$^7$ s and  an average  luminosity of $3 \cdot 10^{30}$ cm$^{-2}$ s$^{-1}$) , 
statistics per LHC run will be $\sim 2.5 \cdot 10^6$ $J/\psi$, and 
$\sim 27000$ $\Upsilon(1S)$. 
During the first 10 years of LHC running it is planned to deliver deuterium lead collisions at $\sqrt{s}$=6.2A TeV 
and the cumulated statistic per d+Pb LHC run will be $\sim 1.5 \cdot 10^6$ $J/\psi$'s and $\sim 15000$ $\Upsilon(1S)$'s 
\footnote{We have not taken into account in this estimation the fact that the center-of-mass frame in d+Pb collisions will present a
relative rapidity of $\pm$0.12 (depending on the deuterium or lead beam direction: d+Pb or Pb+d) 
with respect the laboratory frame due to the asymmetry in the deuterium and lead beam energies.
This will slightly change the acceptance of the Muon spectrometer and could slightly increase or decrease the reported statistics. 
Since we are interested to measure shadowing for small $x$ values inside the lead nucleus, 
the lead beam will come from the back of the spectrometer, and the spectrometer will cover the rapidity range $-3.88<y<-2.28$.
This effect would be more pronounced in p+Pb collisions.}

\subsection{Open heavy flavor measurements}
The measurement of open heavy flavors will also be a priority of our physics program.
Being very interesting in its own, it will allow for a normalization of the quarkonia production rates.
Different alternative analysis will be applied, which exploit the muon production from open charm and beauty mesons via 
their semi-leptonic decay.
\begin{figure}
\begin{center}
\includegraphics[width=8cm,clip]{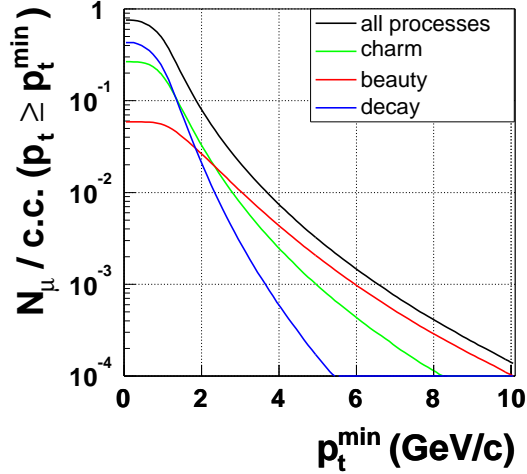}
\caption{\small{Integrated single muon $p_T$ distribution in most central Pb+Pb collisions at 5.5A TeV \protect\cite{Guer04}. }}
\label{fig:SingleMuons}
\end{center}
\end{figure}

\begin{figure}
\begin{center}
\includegraphics[width=11cm,clip]{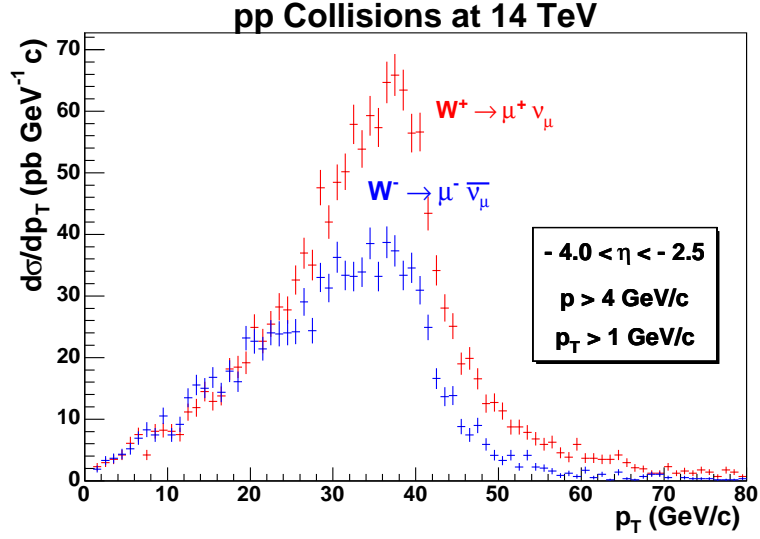}
\caption{\small{Single muon $p_T$ distribution 
coming from the W muon decay in the Muon spectrometer acceptance for p+p collisions at 14 TeV \protect\cite{Cone05}.}}
\label{fig:WSingleMuons}
\end{center}
\end{figure}

\paragraph{Single muon measurements.} 
The measurement of muon p$_T$ distributions will provide the first measurement of heavy quark 
production at high rapidities at LHC. 
For p$_T$ larger than 5-8 GeV/c muon production will be dominated by the semi-muonic decay of D and B mesons. 
The expected integrated single muon contribution in the most central Pb+Pb collisions at 5.5A TeV
is presented in Fig. \ref{fig:SingleMuons}.
For very high $p_T$ a noticeable contribution to the single muon $p_T$ distribution will be originated from 
the muonic decay of the W-bosons. This contribution will be maximum in the $p_T$ range between 30 and 40 GeV/c.
In Fig. \ref{fig:WSingleMuons} the differential cross-section for the muonic channel in the Muon spectrometer acceptance is represented.
During a p+p run at 14 TeV it is expected to detect around 50000 muons coming from the decay of the W boson. 

\paragraph{Muon pairs measurements.} 
Like and unlike sign muon pairs originating from the same hard scattering or same heavy quark, will present a residual 
correlation in the low (1-3 GeV/c$^2$) and high (4-8 GeV/c$^2$) invariant mass ($M_{inv}$) regions. 
Correlated unlike-sign muon pairs in high $M_{inv}$ region will be mainly produced
by semi-muonic decays of $D-\bar{D}$ and  $B-\bar{B}$ mesons from the same hard scattering: 

\begin{picture}(400,150)
\put(0,    75.0){Hadron collision $\Longrightarrow$  $c$ + $~\bar{c}$ + ...}
\put(150, 105){$D^{\circ}$ (or $D^+$, $D^\ast$, $D^{\ast +}$, $D_s^+$)   + ...}
\put(150,  45){$\bar{D}^{\circ}$ (or $D^-$, $\bar{D}^\ast$, $D^{\ast -}$ $D_s^-$) + ... }
\put(320, 135){$\mu^+$ + ...}
\put(320 , 15){$\mu^-$ + ...} 
\put(105,  82){\line(0, 1){24}}
\put(105, 106){\vector(1, 0){35}}
\put(130,  67){\line(0, -1){20}}
\put(130,  47){\vector(1,0){15}}
\put(175, 120){\line(0, 1){18}}
\put(175, 138){\vector(1, 0){135}}
\put(175,  40){\line(0, -1){23}}
\put(175,  17){\vector(1, 0){135}}
\put(325, 127){\vector(0,-1){35}}
\put(325,  22){\vector(0, 1){35}}
\put(270,  75){Residual $\mu^+\mu^-$ correlation $D-\bar{D}$}
\end{picture}

\begin{picture}(400,150)
\put(0,    75.0){Hadron collision $\Longrightarrow$  $b$ + $~\bar{b}$ + ...}
\put(150, 105){$\bar{B}^{\circ}$ (or $B^-$, $\bar{B}^\ast$, $\bar{B}_s^\circ$, ...)   + ...}
\put(150,  45){$B^{\circ}$ (or $B^+$, $B^\ast$, $B_s^\circ$, ...) + ... }
\put(320, 135){$\mu^-$ + ...}
\put(320 , 15){$\mu^+$ + ...} 
\put(105,  84){\line(0, 1){22}}
\put(105, 106){\vector(1, 0){35}}
\put(130,  67){\line(0, -1){20}}
\put(130,  47){\vector(1,0){15}}
\put(175, 120){\line(0, 1){18}}
\put(175, 138){\vector(1, 0){135}}
\put(175,  40){\line(0, -1){23}}
\put(175,  17){\vector(1, 0){135}}
\put(325, 127){\vector(0,-1){35}}
\put(325,  22){\vector(0, 1){35}}
\put(270,  75){Residual $\mu^+\mu^-$ correlation $B-\bar{B}$}
\end{picture}

The low $M_{inv}$ region will be populated by $B-D$ or $\bar{B}-\bar{D}$ semi-muonic decays from the same bottom quark fragmentation.
Like-sign muon pairs will be dominated by $B-\bar{D}$ and, more exotic mechanism, like 
$B^0-\bar{B}^0$ oscillations.
The study of such correlations will allow for measuring the beauty production cross-section \cite{Guer04} and 
to study the relative contribution of NLO diagrams to the total beauty production cross-section.
In HIC, the $\mu^+\mu^-$ residual correlations could be modified by final state effects.
In this respect,  $\mu^+\mu^-$ residual correlations could become a  probe of the heavy quark re-scattering in the QGP medium.

\paragraph{Correlation with the central barrel.}
In the same way, residual correlation between unlike sign muons and electrons (or Kaons) can be exploited to get similar physics informations.
Since electrons and kaons are detected in the ALICE central barrel, such a correlation will be sensible to heavy quark production 
in an intermediate rapidity range $-1.0<y<-2.5$.
Additionally, unlike-sign $e-\mu$  pairs cannot not be generated by any resonance, nor thermal or direct production \cite{Croc05}.

\paragraph{Multi-correlation of muons.} 
Residual correlation of 3 (or more) muons will only be sensible to beauty production.
In particular, 3 muon correlation will probe the beauty production decaying in the charmonium channel which will noticeably contribute 
to the $J/\psi$ total production cross-section.
Unfortunately, the combinatorial background of 3-muon correlations in Pb+Pb central collisions will be huge 
and this technique could only be efficiently applied  in p+p, d+Pb  and (perhaps) in Ar+Ar collisions \cite{Mors05}.

\section{Conclusions and Perspectives}
In summary, the muon spectrometer supports an ambitious physics program in the ALICE experiment, focused on heavy quark physics 
for studying QGP at the LHC collider.
The muon spectrometer has entered in its construction phase and will be ready in 2007 at the interaction point 2 of the LHC
for the first data taking.

\section{Acknowledgments}
I would like to thank my colleagues of ALICE collaboration for fruitful discussions.
In particular, I would like to thank Philippe Crochet, Christian Finck, Rachid Guernane, Smbat Grigoryan and Andreas Morsch 
for helping me in the preparation of this talk. 
I would like to thank Laurent Aphecetche, Zaida Conesa del Valle, Philippe Crochet and Christian Finck for carefully reading this manuscript.

\end{document}